\documentclass[twocolumn,twoside,preprintnumbers,amsmath,amssymb,showkeys]{revtex4}
\usepackage{epsfig}
\usepackage{graphicx}

\usepackage{fancyhdr}

\usepackage{pslatex}

\begin{document}

\title{$p_T$ distribution of hyperons in $200A$ GeV Au-Au in smoothed particle hydrodynamics}

\author{Wei-Liang Qian$^1$, Rone Andrade$^1$, Ot\'{a}vio Socolowski Jr.$^3$, Fr\'{e}d\'{e}rique Grassi$^1$, Takeshi Kodama$^2$ and Yogiro Hama$^1$}

\affiliation{
$^1$ Instituto de F\'{\i}sica, Universidade de S\~ao Paulo, C.P. 66318, 05315-970 S\~ao Paulo-SP, Brazil \\
$^2$ Instituto de F\'{\i}sica, Universidade Federal do Rio de Janeiro, C.P. 68528, 21945-970, Rio de Janeiro, Brazil\\
$^3$ Departamento de F\'{\i}sica, Instituto Tecnol\'{o}gico da  Aeron\'{a}utica, C.P. 12228-900, S\~ao Jos\'e dos Camos, SP, Brazil}

\received{on October 31, 2006; revised version received on April 17, 2007}

\begin{abstract}
The transverse momentum distributions of hadrons in 200GeV Au-Au collisions at RHIC are calculated using the smoothed particle hydrodynamics code SPheRIO. They are compared with data from STAR and PHOBOS Collaborations. By employing an equation of state which explicitly incorporates the strangeness conservation and introducing strangeness chemical potential into the code, the transverse spectra give a reasonable description of experimental data,
except the multiplicities of hyperons.


\end{abstract}
\maketitle

\thispagestyle{fancy}

\setcounter{page}{1}

\section{Smoothed particle hydrodynamic model}
The hydrodynamic description of high-energy nuclear collisions was
first introduced by L.D Landau \cite{landau}. Ever since then, 
the model has successfully accounted
for certain features in nuclear collisions \cite{clare,stocker}, such as
the energy dependence of the average multiplicity and the transverse-momentum
distributions. The hydrodynamic
model employs the hypothesis that hot and dense matter, formed 
in high energy nuclear collisions,
reaches local thermal equilibrium, after which it expands and cools down
before particle emission takes place. The local thermal equilibrium state
serves as the initial condition for the hydrodynamic model,
and it is expressed in terms of distributions of the fluid
velocity and of thermodynamical quantities for a given time-like parameter. 
The hydrodynamical expansion is described by the
conservation equations of the energy-momentum, baryon number, and other
conserved charges, such as strangeness, isospin, etc. In order to close
the system of partial differential equations, one also needs the equation of state (EOS)
of the fluid. With the fluid being cooled and rarefied further,
the constituent particles will finally reach the stage where they do not
interact with each other until they reach the detector, namely, the
decoupling stage of the hydrodynamic model.

It can be shown, that the
whole set of relativistic hydrodynamic equations can be derived by using the
variational principle, taking matter flow as the variable \cite{elze}.
However, the resulting system of equations is highly nonlinear,
and since, in general, there is no symmetry in the system,
the direct integration of these equations is very expensive from the computational point of view.
The aim of the smoothed particle hydrodynamic (SPH) algorithm  
for nucleus-nucleus collision is to provide a rather simple 
scheme for solving the hydrodynamic equations. Such method does 
not have to be too precise, but rather robust enough to deal 
with any kind of geometrical structure. The SPH
algorithm was first introduced for astrophysical applications \cite{astro}. It
parametrizes the matter flow in terms of discrete Lagrangian coordinates, namely,
of SPH particles. In the model, usually some conserved quantities, such as
baryon number and entropy are assigned to them.
In terms of SPH particles, the hydrodynamic equations
are reduced to a system of coupled ordinary differential equations.

The code which
implements the entropy representation of the SPH model for relativistic high
energy collisions \cite{va}, and which has been investigated and developed within
the S\~{a}o Paulo - Rio de Janeiro Collaboration, is
called SPheRIO. As it has been shown, the entropy representation of the SPH
model is an efficient and practical method to tackle the problems concerning
relativistic high-energy nucleus-nucleus collisions, which are characterized
by highly asymmetrical configurations. This method has successfully been used
to investigate the effects of the initial-condition fluctuations and adopting
the continuous emission scenario for the description of decoupling process 
\cite{va,topics,ce,ebe,strange}.

In this work, we focus on the role of strangeness and study
its effects on hydrodynamic models for high-energy nucleus-nucleus collisions.
The SPH method is employed in our calculation of the
$p_T$ spectra of hyperons in $200A$ GeV Au-Au collisions
and the results are compared with experimental data.
The strangeness enters in the model in two different ways. Firstly, 
in the calculation, a different set of EOS
is employed while strangeness conservation is considered.
The new set of EOS is expected to have an impact on the hydrodynamic evolution.
Secondly, strangeness chemical potential is incorporated in the model for
hyperons and strange mesons, while calculating their spectra.
Therefore the multiplicity for strange hadrons will be modified 
because of the introduction of strangeness chemical potential.
As in ref \cite{topics}, we use the hadronic resonance model with finite
volume correction to describe the matter on the hadronic side, where the
main part of observed resonances in Particle Data Tables has been included.
For quark gluon plasma (QGP) phase, the ideal gas model is adopted.

\section{Results and discussion}

We illustrate in Fig.~1, the phase boundaries for the EOS we used in the
following calculation considering strangeness in terms of temperature as
a function of baryon density.
In Fig.~2 the phase boundary is depicted in a plot of temperature as a function of 
strangeness chemical potential. It is worth noting here that due to the 
introduction of strangeness conservation,
the strangeness chemical potential does not remain constant during the isothermal compression
procedure.

\begin{figure}[!htb]
\vspace*{-1.1cm}
\begin{center}
\includegraphics*[width=9.cm]{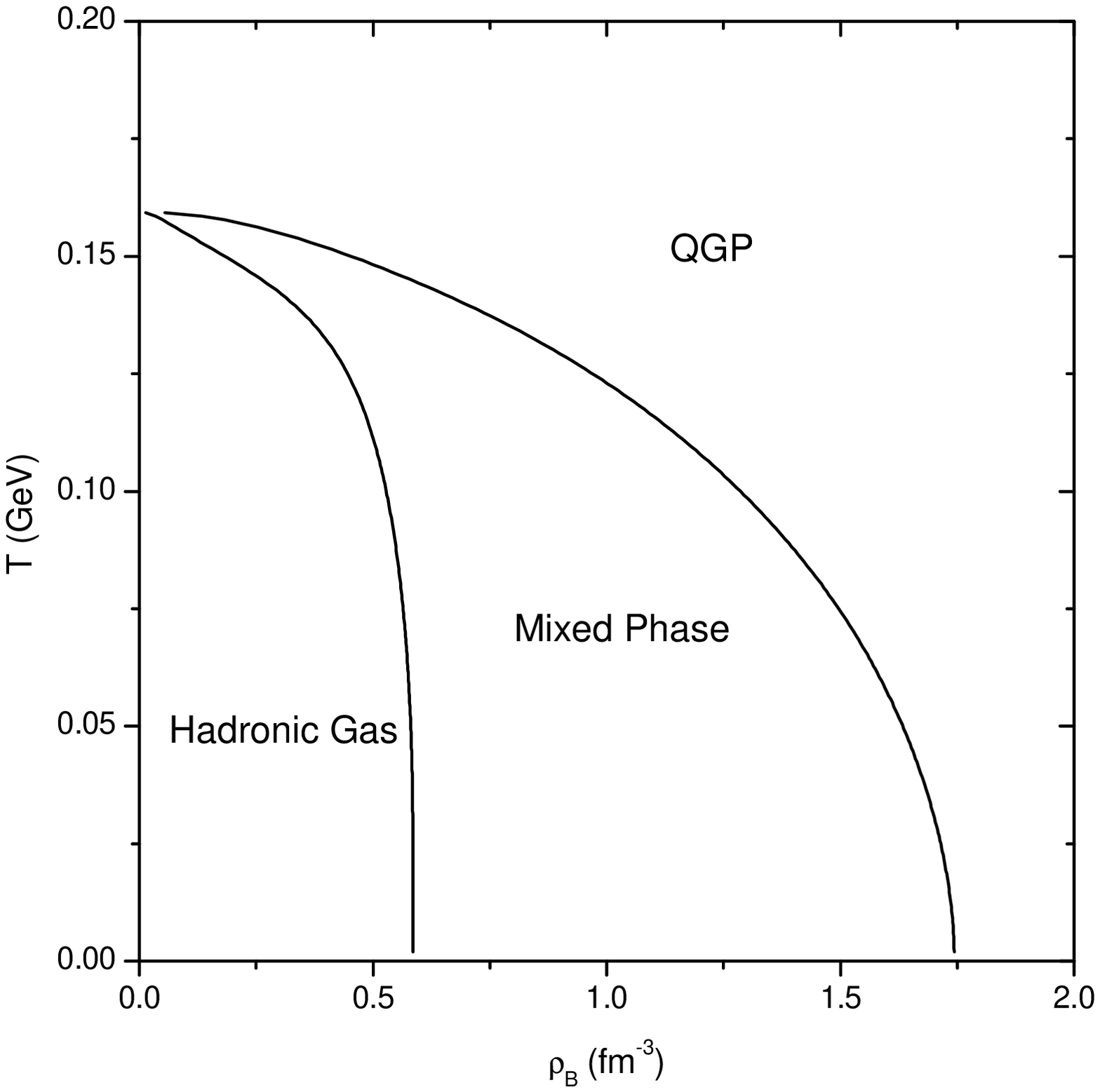} 
\end{center}
\vspace*{-5.cm}
\caption{Phase boundary for EOS with strangeness,
depicted in the temperature vs. baryon density plane.} 
\label{figure:fig1}
\end{figure} 

\begin{figure}[!htb]
\vspace*{-1.1cm}
\begin{center}
\includegraphics*[width=9.cm]{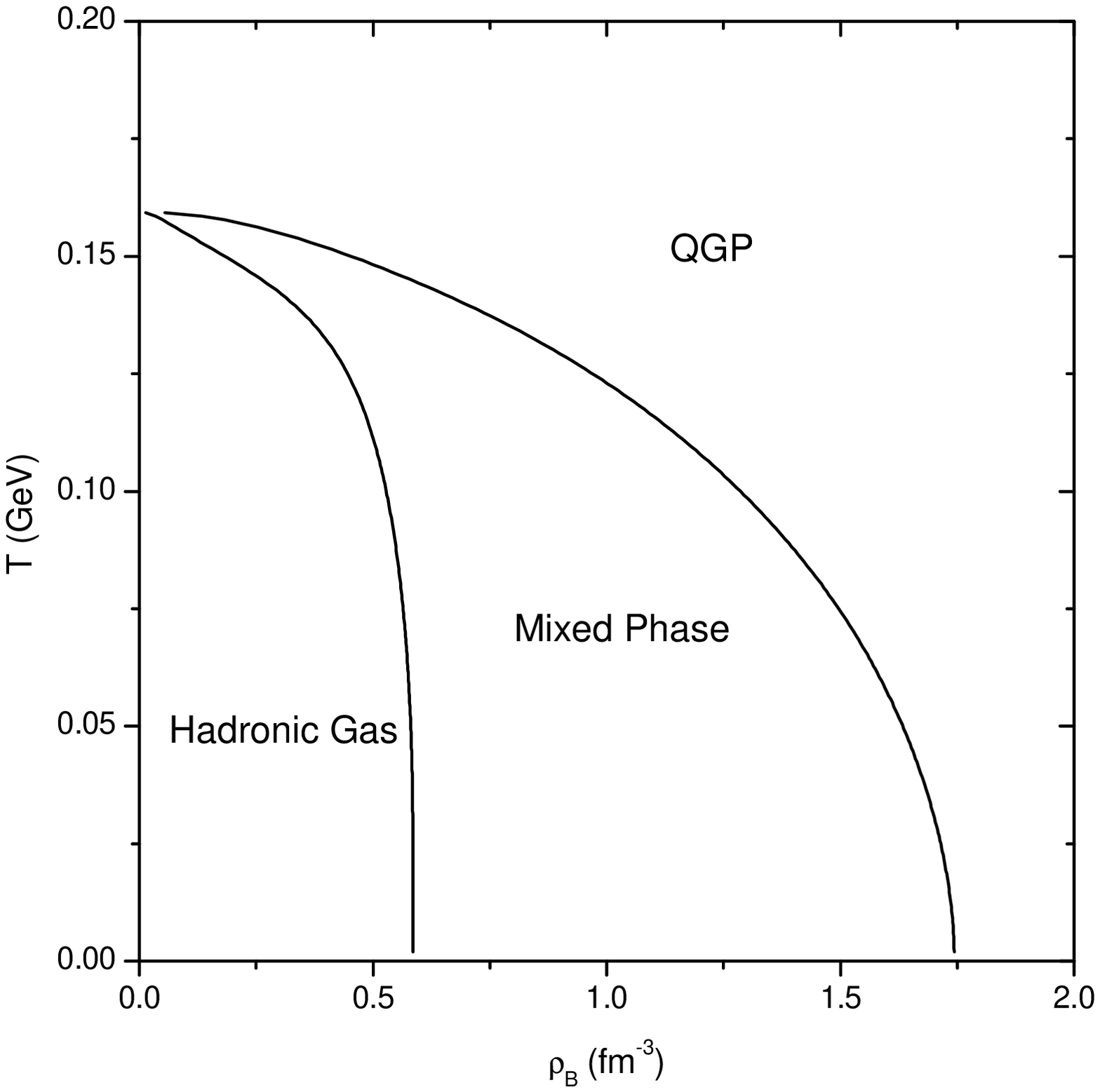} 
\end{center}
\vspace*{-5.cm} 
\caption{Phase boundary for EOS with strangeness,
depicted in the temperature vs. strangeness chemical potential plane.} 
\label{figure:fig2}
\end{figure} 

As in the previous works \cite{va,topics,ce,ebe,strange},
we use the NEXUS event generator to produce the initial conditions.
However, we make use of a rescaling factor to fix
the pseudo-rapidity distribution of all charged particles.
Furthermore, following ref.~\cite{peter}, here we introduce an
additional transverse velocity, which was needed for the present
choice of parameters.
Thus the initial transverse velocity reads
\begin{eqnarray}
v_T &=& NeXUS + \tanh(\alpha r)\hat{r}
\end{eqnarray}
where $r$ is the radial distance from origin, and $\alpha$ is a small parameter. We also assume that 
the strangeness density is locally zero everywhere in the system. 

As a preliminary application, we have calculated the
pseudo-rapidity distribution of all charged particles. 
\begin{figure}[!htb]
\vspace*{-1.cm}
\begin{center}
\includegraphics*[width=8.2cm]{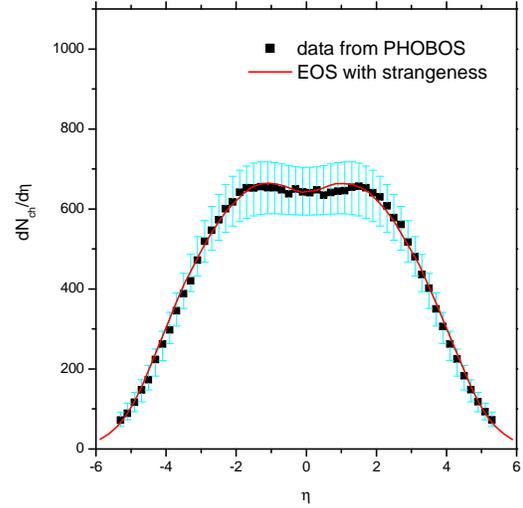} 
\end{center}
\vspace*{-4.5cm}
\caption{The pseudo-rapidity distribution of all charged particles for the
most central Au-Au collisions at 200A GeV. The data are from the PHOBOS Collab.} 
\vspace*{-.2cm} 
\label{figure:fig3}
\end{figure} 

In Fig.~3, we
present the result for the most central Au-Au collisions
at 200A GeV. The experimental
data are from PHOBOS Collaborations, taken in the most central Au-Au at 200A GeV.
The experimental transverse momentum distribution data of all charged
particles can be well reproduced with
a choice of $\alpha=0.04 $fm$^{-1}$ and freeze out temperature $T_f = 110$ MeV.
In Fig.~4 the transverse momentum
distributions with freeze out temperatures 110 MeV 
are depicted, together with the experimental data. The experimental
data are taken from the STAR Collaboration \cite{star1}, taken in the most central Au-Au at 200A GeV,
with $\eta=0$. 
\begin{figure}[!bht] 
\vspace*{-1.1cm}
\begin{center}
\includegraphics*[width=7.6cm]{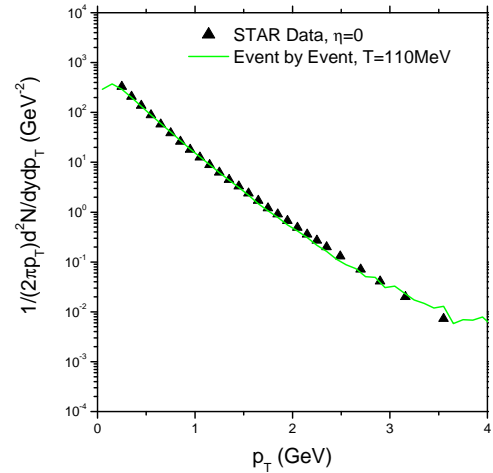} 
\end{center}
\vspace*{-4.3cm}
\caption{Transverse momentum distribution of all charged particles for the
most central Au-Au collision at 200A GeV with rapidity interval $-1.0 < \eta <
1.0$. The data are from the STAR Collab.} 
\label{figure:fig4}
\end{figure} 

In Figs.~5 and 6, we use the parameters fixed above 
to calculate the spectra of other hadrons. In Fig.~5 
we show the transverse mass spectra of pion, proton and 
kaon for the most central collisions at mid rapidity, 
as well as the experimental\hfilneg\ 

\newpage 
\noindent data from the BRAHMS  Collaboration \cite{brahms1}.
In Fig.~6, the spectra of $\Lambda$, $\Xi$ and $\Omega$ are depicted,
together with the data from the STAR Collaboration \cite{star2}. 
It is found that the present hydrodynamic model gives a good
description of the transverse momentum spectra of light hadrons as pion, kaon and proton.
As for hyperons which are much heavier,
it gives a reasonable description of the shape of the curves,
while there are visible discrepancies in the multiplicities.
\begin{figure}[!htb]
\vspace*{-.5cm}
\begin{center}
\includegraphics*[width=7.6cm]{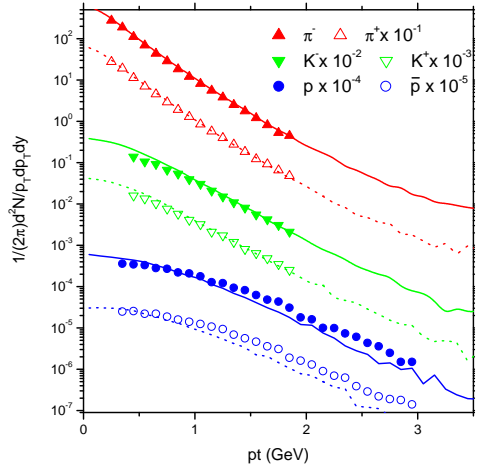} 
\end{center}
\vspace*{-4.cm}
\caption{Transverse momentum distribution of pion, kaon and proton for the
most central Au-Au collisions at 200A GeV with rapidity interval $-1.0 < \eta <
1.0$. The data are from the BRAHMS Collab.} 
\label{figure:fig5}
\end{figure} 
This may indicate that the system undergoes first the chemical freeze-out 
at high temperature where the observed particle multiplicities are fixed 
and next the thermal freezeout at lower temperature where the shape of 
the transverse momentum distribution is fixed \cite{chfz0}. The present 
results of hyperon spectra might be improved by introducing the chemical 
freeze-out mechanism into the model \cite{chfz1,chfz2,chfz3}. Such work is 
under progress. 

It is inferred from the comparison with the EOS without 
strangeness that the difference between the two sets of EOS at 
mid-rapidity region is not very large. However, as it was\hfilneg\  

\newpage 

\begin{figure}[!thb]
\vspace*{-.9cm}
\begin{center}
\includegraphics*[width=7.8cm]{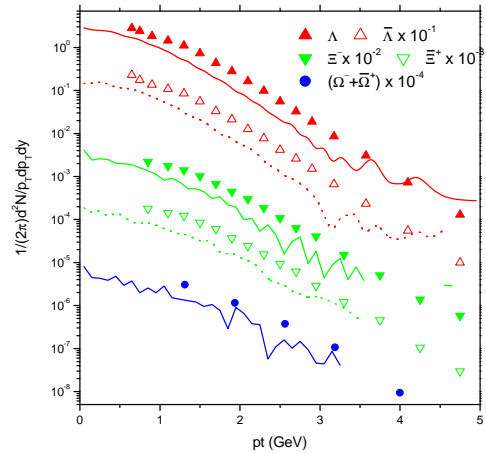} 
\end{center}
\vspace*{-4.5cm}
\caption{Transverse momentum distribution of $\Lambda$s, $\Xi$s and $\Omega$s for the
most central Au-Au collisions at 200A GeV with rapidity interval $-1.0 < \eta <
1.0$. The data are from the STAR Collab.} 
\label{figure:fig6}
\end{figure} 

\noindent indicated in Ref.~\cite{kdis} for Pb-Pb  
collisions at 158A GeV, as one goes to the high-rapidity region, the difference is expected to become much more significant. 
We hope similar data also for Au-Au collisions at RHIC be obtained 
in near future. 

\vspace*{1.cm} 

\noindent{\bf Acknowledgements} 
\medskip 

This work has been supported by FAPASP under the contract numbers 2004/10619-9, 2005/54595-9, 
2004/13309-0 and 2004/15560-2, CAPES, CNPq, FAPERJ and PRONEX.


\begin{thebibliography}{99}

\bibitem{landau}L.D. Landau, Izv. Akad. Nauk SSSR {\bf17} (1953) 51;
S.Z. Belenkij and L.D. Landau, Usp. Fiz. Nauk {\bf56} (1955) 309
\bibitem{clare}R.B. Clare and D. Strottman, Phys. Rep. {\bf141} (1986) 177
\bibitem{stocker}H. St${\rm\ddot o}$cker and W. Greiner, Phys. Rep. {\bf137} (1986) 277
\bibitem{elze}H. Elze, Y. Hama, T. Kodama, M. Makler and J. Rafelski, J. Phys. {\bf G25} (1999) 1935
\bibitem{astro}L.B. Lucy Astrophys. J. {\bf82} (1977) 1013;
R.A. Gingold and J.J. Monaghan, Mon. Not. R. Astro. Soc. {\bf181} (1977) 375
\bibitem{va}C.E. Aguiar, T. Kodama, T. Osada, Y. Hama, J. Phys. G{\bf 27} (2001) 75;
\bibitem{topics}Y. Hama, T. Kodama and O. Socolowski Jr., Braz. J. Phys. {\bf35} (2005) 24
\bibitem{ce}F. Grassi, Braz. J. Phys. {\bf35} (2005) 52
\bibitem{ebe}O. Socolowski Jr., F. Grassi, Y. Hama, T. Kodama, Phys. Rev. Lett. {\bf93} (2004) 182301
\bibitem{strange}F. Grassi, O. Socolowski, Jr., Phys. Rev. Lett. {\bf 80} (1998) 1170
\bibitem{peter}P.K. Kolb and R. Rapp, Phys. Rev. {\bf C67} (2003) 044903
\bibitem{star1}STAR Collab., Phys. Rev. Lett. {\bf 91} (2003) 172302
\bibitem{brahms1}BRAHMS Collab., Phys. Rev. C{\bf 72} (2005) 014908
\bibitem{star2}STAR Collab., nucl-ex/0606014
\bibitem{chfz0}E.V. Shuryak, Nucl. Phys. A{\bf 661} (1999) 119;
U. Heinz, ibid. A{\bf 661} (1999) 140
\bibitem{chfz1}T. Hirano and K. Tsuda, Phys. Rev C{\bf 66} (2002) 054905
\bibitem{chfz2}D. Teaney, J. Lauret and E.V. Shuryak, nucl-th/0110037;
D. Teaney, nucl-th/0204023
\bibitem{chfz3}C. Nonaka and S. A. Bass, Phys. Rev. C{\bf 75} (2007) 014902
\bibitem{kdis}NA49 Collab., J. Phys. G{\bf 23} (1997) 1817

\end{thebibliography}
\end{document}